# HI signatures of galaxy evolution

**J.M. van der Hulst**[1]
*Kapteyn Astronomical Institute*
*Landleven 12, NL-9747 AD, Groningen, The Netherlands*
*E-mail:* `j.m.van.der.hulst@rug.nl`

HI in and around galaxies provides unique information about the various processes shaping galaxies: merging, cold gas accretion, feedback. Observations of galaxies in the nearby universe are beginning to reveal the HI signatures of these processes by pushing the sensitivity limits of existing radio synthesis telescopes to their limits. This paper gives a brief inventory of these signatures. The capabilities of new instruments such as the SKA pathfinders and precursors is briefly addressed, though ultimately SKA will provide the adequate sensitivity to find these HI signatures beyond the local universe.

[1]  Speaker

## 1. Introduction

In the past four decades many hundreds of galaxies have been observed in detail in HI with radio telescopes such as the Westerbork Synthesis Radio Telescope (WSRT), the Very Large Array (VLA), the Australia Telescope Compact Array (ATCA) and the Giant Metrewave Radio Telescope (GMRT). Over the past decade the increased sensitivity has revealed a lot of information about HI at low column density levels ($< 10^{20}$ cm$^{-2}$) in and around galaxies.

While the original quest of HI projects was to obtain detailed distributions and velocity fields, in particular to determine the dark matter content of galaxies, the emphasis has now shifted to using the HI to look for signatures that tell us about the process of galaxy formation, the formation of stars and the effects of recent star formation on the immediate environment. A detailed account of many interesting examples of such signatures can be found in a recent review paper [19].

Here I will briefly summarise some of these results and discuss the potential of future instruments such as the Square Kilometre Array (SKA) and its pathfinders and precursors.

## 2. HI in and around galaxies in the nearby universe

Sensitive HI observations of nearby galaxies have provided a number of interesting phenomena which can be connected to the complex process of galaxy evolution. These phenomena (e.g. see [19]) can be categorised as follows: warped and/or asymmetric HI disks, disks with kinematic asymmetries, extra-planar HI, and HI with anomalous velocities. The former two phenomena are global features, where the latter two features are often restricted to the inner, star forming disk. Here I briefly summarise the evidence.

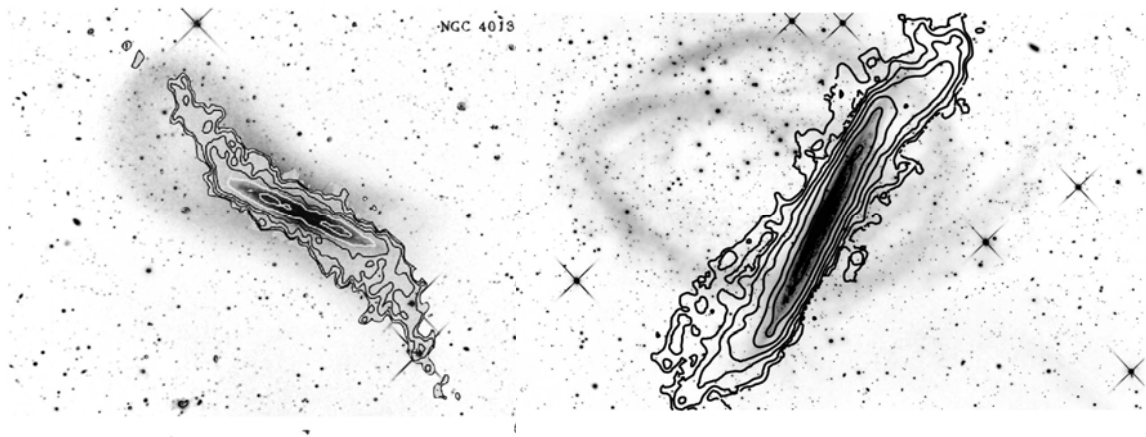

**Figure 1.** Warped HI column density distributions overlaid on deep optical images illustrating the presence of stellar streams [4,10,11,21]. NGC 4013 is shown in the left panel, NGC 5907 is shown in the right panel.

Warps were already discovered in the early days of HI synthesis observations [16]. The frequency of warps is surprisingly high ( > 50%, e.g. see [19] for a summary) and their originin still is not well understood. The discovery of faint stellar streams in NGC 4013 and NGC 5907, shown in Figure 1 with the HI column density distributions overlaid [4,10,11,21], suggests a connection with the accretion of material from small satellites.  Warps are not only apparent in edge-on galaxies, where they are easy to detect, but can also be found in more inclined galaxies from an analysis of the velocity field under the assumption that the HI is moving in circular orbits about the centre of mass of the galaxy. A nice example of such a galaxy is NGC 5055 [1], shown in Figure 2. What is remarkable about this galaxy is that the outer, warped disk is fairly symmetric, exhibits spiral structure and star formation associated with the brightest HI patches (as indicated by the GALEX UV image) and that the kinematics of the outer part indicated that the possibility exists that the center of mass of the bright inner disk is not coincident with the center of mass of the dark halo. So in addition to the precise structure and geometry of the warp such observations also provide information about the dynamical state of the different mass components.

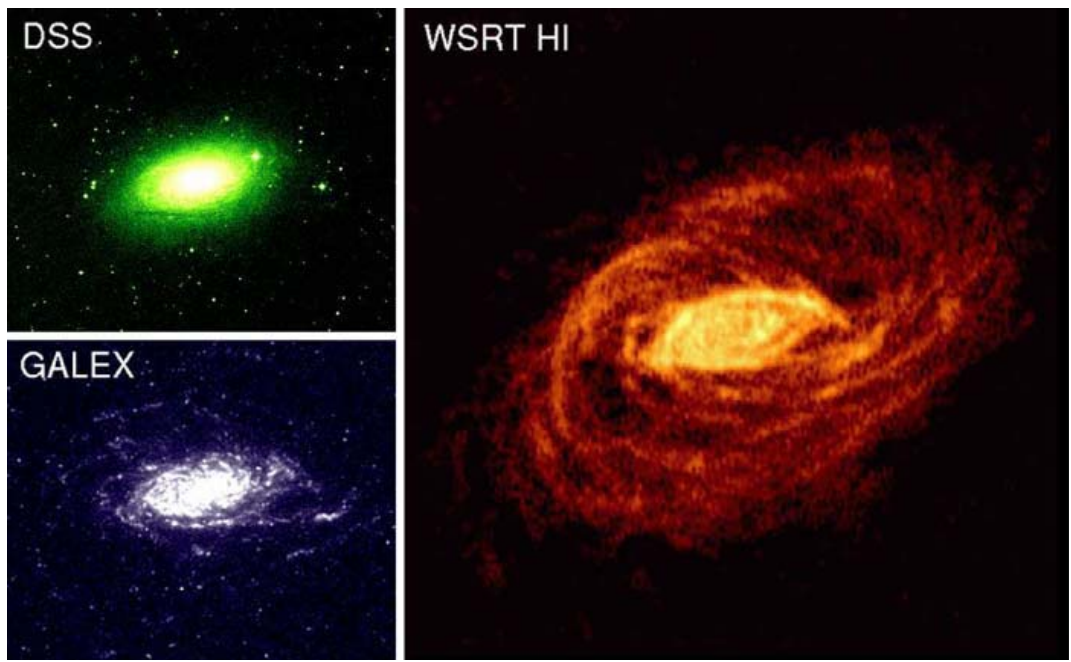

**Figure 2.** Optical (*DSS*), GALEX and WSRT HI images of the warped galaxy NGC5055 (all on the same scale). Column densities range from about $3 \times 10^{19}$ cm$^{-2}$ to $1 \times 10^{21}$ cm$^{-2}$ [1].

Many of the features other than warps are present in the galaxy M 101 [9,25]: lopsidedness (both in the HI distribution and the kinematics), a large HI complex of 2 x 10$^8$ M$_\odot$, presumably extra-planar, moving with velocities up to 150 km s$^{-1}$ with respect to HI in the local disk, and anomalous velocity gas in the inner disk, with as most notable example the symmetric super bubble in NGC 5462 [9] (see Figure 3).

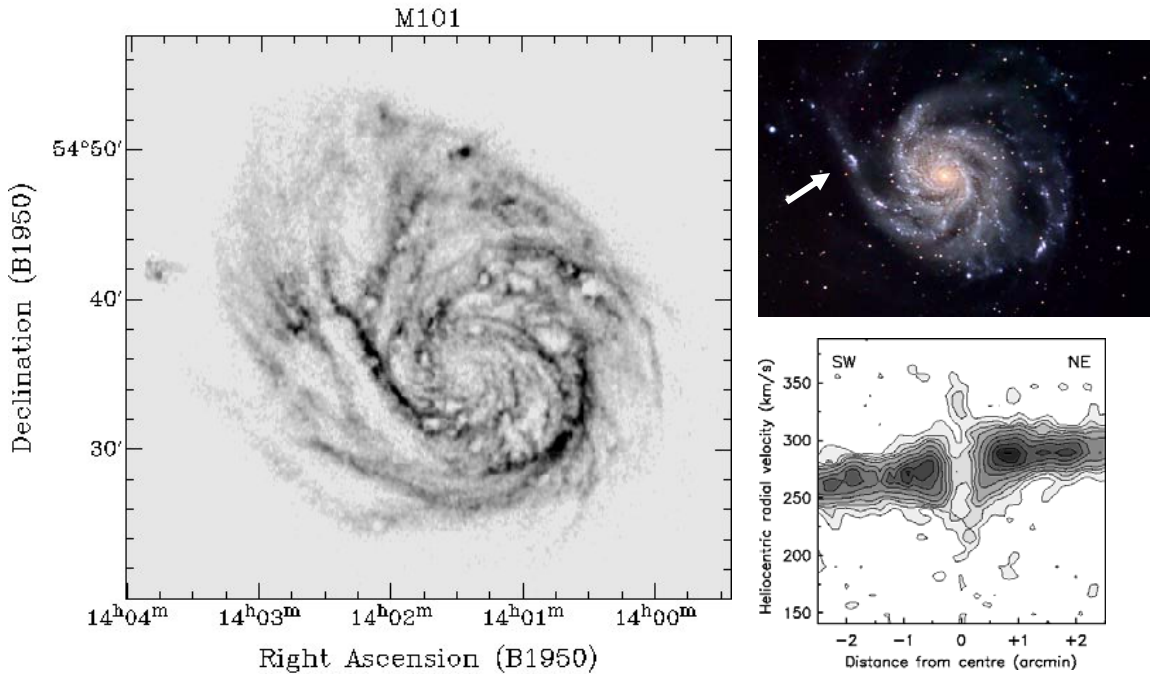

**Figure 3.** Lopsided HI distribution in M 101 (left panel) and optical image (top right panel) on the same scale. The lower right panel shows a position velocity diagram [9] centered at the HII region NGC 5462 (see arrow in the top right panel) indicating the symmetric HI outflow around the HI hole in NGC 5462.

Anomalous velocity gas has subsequently been found in many galaxies and with varying signatures. In NGC 6946 anomalous velocity gas appeared to be widespread throughout the star forming disk [2], both at positive and negative velocities with respect to the local rotation. In the inner disk the presence of holes and this anomalous velocity gas can be related to the active star formation suggesting outflows of gas as a result of supernovae and stellar winds. Figure 4 illustrates how widespread this anomalous velocity gas is. It shows the kinematics of the HI in a velocity position diagram of the entire galaxy's emission integrated along the minor axis with the effects of rotation taken out. The gas 'above' and 'below' the central HI disk clearly demonstrates the anomalous velocity gas, presumably outflows. It is very clear that the bulk of the anomalous velocity gas is present in the star forming disk, as outlined by the Hα emission show at the bottom of Figure 4.

In the outer parts, however, such holes and anomalous velocity gas needs another explanation and can be the result of infalling material. At low column density levels NGC 6946 does display a 20 kpc long plume in the north west containing of $7.5 \times 10^7$ $M_\odot$ of HI (e.g. Figure 5 in [19] ). This HI mass is not dissimilar from that of two satellite galaxies farther to the west ($1.2 \times 10^8$ $M_\odot$ and $8.8 \times 10^7$ $M_\odot$ respectively). This HI plume smoothly blends (also kinematically) into the HI in the disk on NGC 6946. This arrangement is very suggestive for recent accretion of HI (and perhaps other material).

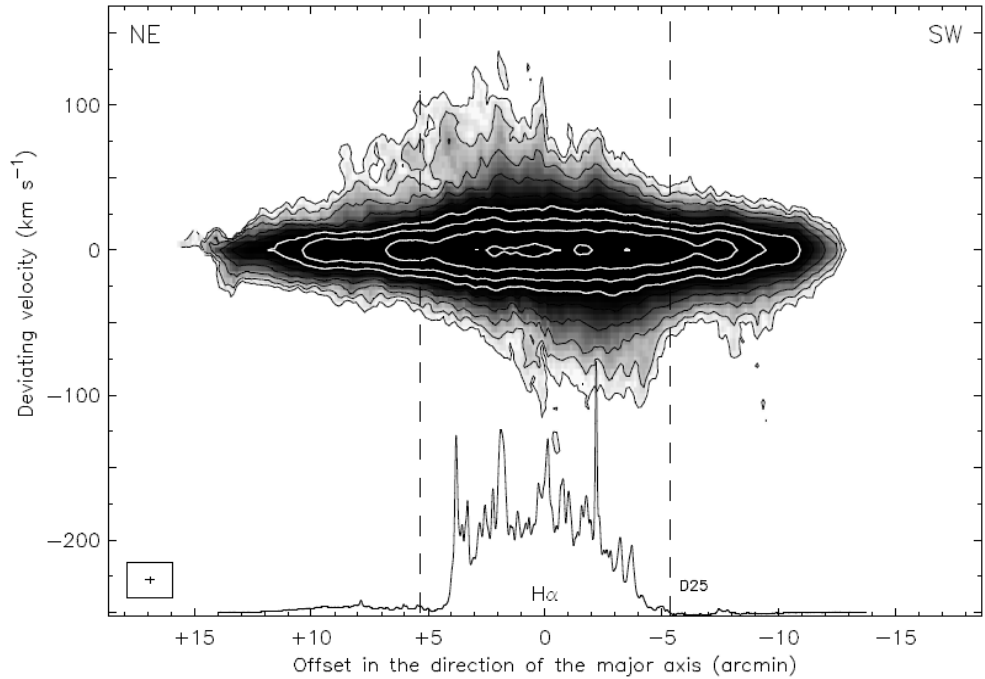

**Figure 4.** A total H I position-velocity diagram for NGC 6946 parallel to the major axis after strip integration of the derotated H I emission along the minor axis. The bottom profile shows the Hα emission, also integrated along the minor axis. The dashed lines indicate the $D_{25}$ of NGC6946 [2].

The widespread anomalous velocity gas its distribution perpendicular to the disk can be examined in edge-on galaxies. This was first done in NGC 891 where indeed gas at low column density levels was found extending to 5 kpc above the disk above the disk [24] and corotating with the HI in the disk, albeit at a velocity lower by ~ 25 km s$^{-1}$. Such a more slowly rotating 'thick' HI disk has also been found in NGC 2403 [6,7]. The combination of the slower rotation and the non-zero velocity component perpendicular to the thin HI disk allow separation of this component from the thin HI disk in inclined galaxies. The origin of this component is not clear. The most likely explanation is that most of this gas originates in the disk and has been brought into the halo by a galactic fountain fed by the star formation in the disk. There are, however, several reasons, including the detailed kinematics, to suspect that part of the extra-planar gas has an external origin. This notion is strengthened by the NW HI plume in NGC 6946 [19], the very extended extra-planar gas in NGC 891 [13] shown in Figure 5 and the general picture of HI extensions and asymmetries in many galaxies (e.g. several examples in [19] ). The long filaments to the north west of NGC 891 extend more than 15 kpc above the plane and both the distribution and kinematics strongly suggest that this is infalling material.

Once the column density sensitivity is pushed to these levels ( $< 10^{19}$ cm$^{-2}$) the picture that emerges for the HI in and around galaxies becomes increasingly more complex than just a simple differentially rotating thin HI disk. Many of the additional features provide information that points at physical processes associated with both the ongoing star formation in the disk and the accretion of new material. Some of the accreted material can have originated in the parent galaxy and have been brought out by either a 'galactic fountain' in the inner disks or by gravitational interaction with satellite galaxies disturbing the outer parts.

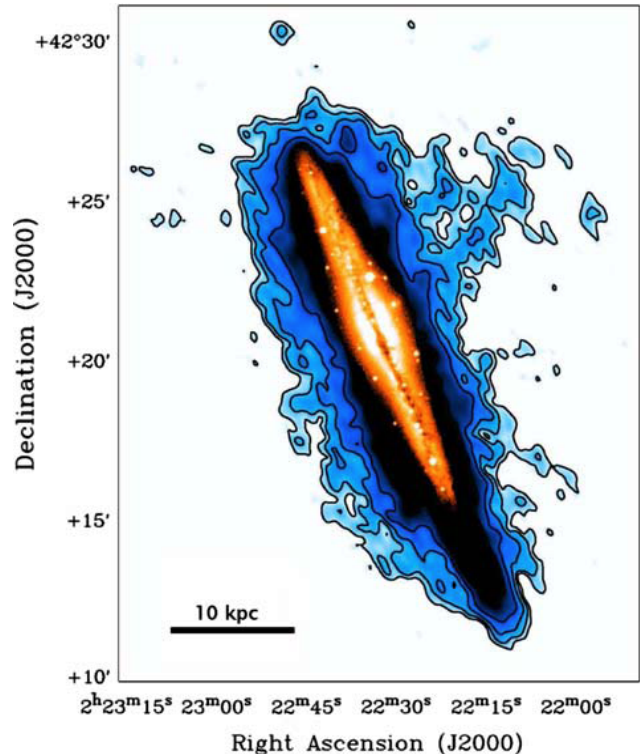

**Figure 5.** Optical DSS image (*red*) and total HI map (*contours + blue shade*) of the edge-on galaxy NGC 891. HI contours are 1, 2, 4, 8, 16 × $10^{19}$ cm$^{-2}$ [13]. The beam size is 25" or 1.1 kpc

This picture strengthens the notion that feedback mechanisms play an important role in galaxy evolution. It is of great interest to note that also from the point of view of theory and simulations the role of of feedback is becoming crucial for matching theory with existing observational facts such as the shape of the HI mass function and the optical luminosity function of galaxies as a function of cosmic time [15,16]. It is clearly important to have a detailed account of how frequent the various HI phenomena described above occur in galaxies in the local universe, so that one can assemble a complete picture of the physics of galaxy formation and evolution and include the observed phenomena in the models and simulations.

In the description so far only spiral galaxies have been used as examples. In recent years it has become very clear that also many Ellipticals and early type galaxies also have detectable amounts of HI [12,20] which can be used to probe phenomena such as the ones described above.

In [19] an attempt is made to estimate from an inventory of some 350 spiral and irregular galaxies, observed in HI with the WSRT as part of the WHISP programme [27], at what rate fresh HI is accreted by galaxies. The estimates are that 0.1 to 0.2 $M_\odot$ yr$^{-1}$ of HI are accounted for by the observations. This is not the total accretion rate as a large fraction of the accreting gas may be ionised and is not seen in HI observations. The star formation rates in galaxies such as NGC 891 and NGC 6946 require about ten times higher accretion rates. If the majority of the gas is ionised, as is the case in the overall cosmic web, then these figures can be brought into agreement. Little is known about the neutral and molecular fractions of the gas in the cosmic web. Recent simulations [17] begin to address this question, which will be very relevant for linking the phenomena described here to the general problem of the physics of galaxy formation and evolution.

### 3. Role of the environment

In the previous section no attention was paid to the possible role of the environment. It has been well established that the dense cluster environment has a definite effect on the size of the HI disks of galaxies, as a result of ram pressure stripping and galaxy-galaxy interactions [5,28]. The most impressive visual representation of these effects is captured in Figure 6 from [5] which displays the HI distributions of galaxies in the Virgo cluster.

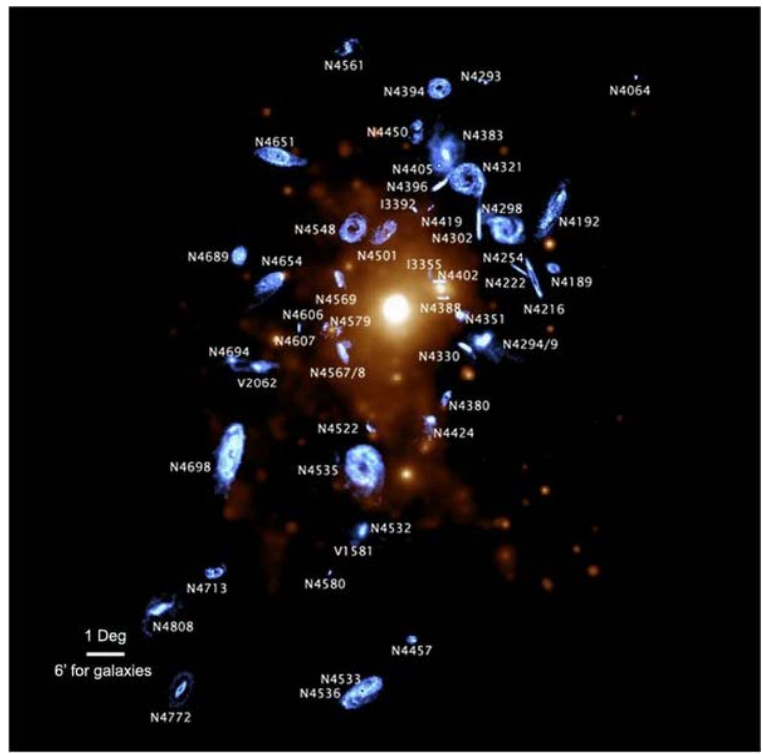

**Figure 6.** Rosat X-ray image (*orange*) of the central part of the Coma cluster with total HI images (*blue*) from [5]. The HI images have been enlarged by a factor 10 to display the details better. Note the variety of morphologies and the trend toward smaller HI disks in the cluster core.

The most extreme contrast to the cluster environment discussed at this conference [28,29] is the void environment. Recently a survey of galaxies in well defined voids, found using the 3D galaxy distribution provided by the SDSS by applying new techniques to determine the density distribution (e.g. see [22,23,28] ) has provided information about the HI properties of galaxies in these truely isolated environments. It appears that despite the pristine and extremely low density environment the HI properties of galaxies are not dissimilar from galaxies of similar morphological type in denser environments. The galaxies in the void environment appear to be a surprisingly interesting collection of perturbed and interacting galaxies, all with small stellar disks [22,23, 28]. Of the first 14 systems analysed in detail four galaxies have significantly perturbed HI disks, five have previously unidentified companions at distances ranging from 50 to 200 kpc, two are in interacting systems, and one was found to have a polar HI disk [22] (Figure 7). Our initial findings suggest void galaxies are a gas-rich, dynamic population which present evidence of ongoing gas accretion, major and minor interactions, and filamentary alignment despite the surrounding under-dense environment.

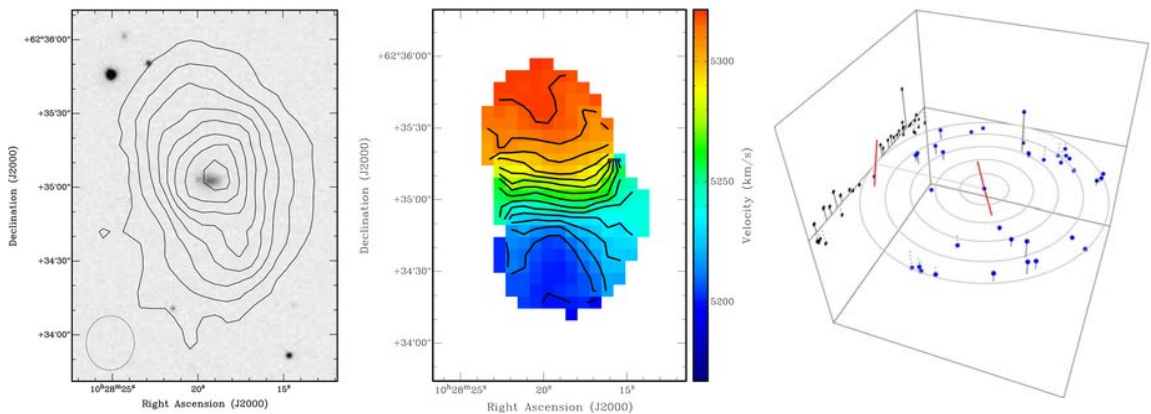

**Figure 7.** SDSS J102819.24+623502.6 from [22]. On the left, a *g*-band image overlaid with HI column density contours of 0.76 (1.9σ), 2.0, 3.3, 4.5, 5.9, 7.0, 8.4, and 9.6 x $10^{20}$ cm$^{-2}$. In the middle, the intensity weighted velocity field overlaid with 8.5 km s$^{-1}$ contours. At the right a 3D overview of the location and orientation of the polar disk (indicated by the red line) within a wall between two voids. The full volume of the sphere with galaxies brighter than $m_g$ = 17.76 out to 10 Mpc has been plotted, with concentric circles every 2 Mpc in the plane of the wall. This demonstrates the loneliness of this galaxy and the emptiness of the bounding voids. An edge-on view is projected on the left, showing the thinness of the wall.

That the formation of galaxy disks appear to be a continuing process as witnessed by the HI signatures demonstrated above is a very interesting discovery. Another very telling group of objects is the void galaxy SDSS 131606.19+413004.2 (MCG +07-27-056) and its neigbours which all have HI and display HI tails and bridges. The object to the east of SDSS 131606.19+413004.2, Mrk 1477 (MCG +07-27-057) also happens to have a polar disk/ring as SDSS J102819.24+623502.6 described above [22] and in addition display a large, fairly straight HI tail pointing toward the north east. It is clear that despite the very isolated environment galaxy formation appears to progress similarly as in denser environments, perhaps at a slower pace.

The precise interpretation awaits completion of the HI and complementary optical, near-IR and Hα imaging of the full void galaxy sample of some 66 galaxies. It is already very clear, however, that the HI morphologies and kinematics play an important role in determining the evolutionary state of the galaxies under study.

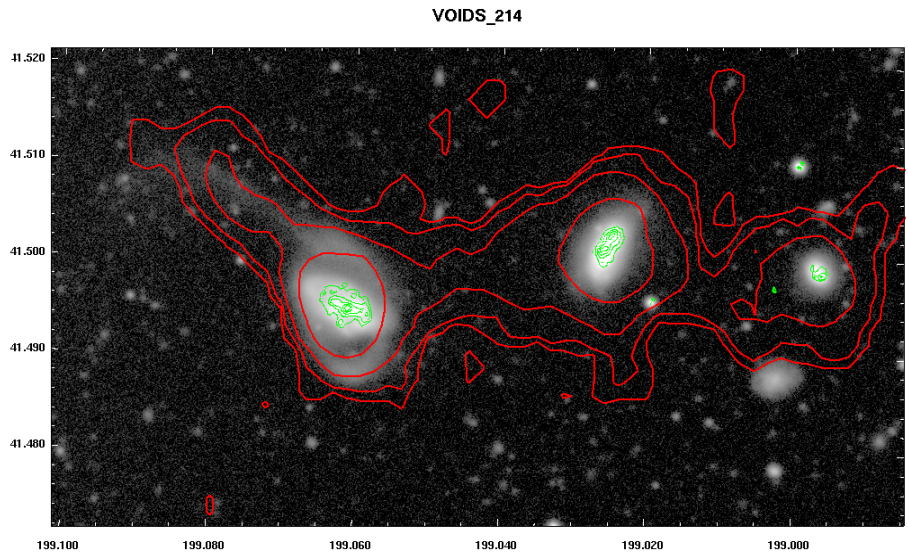

**Figure 8.** The void galaxy SDSS 131606.19+413004.2 (middle object) and neighbouring galaxies. All three are detected in HI (red contours) clearly indicating interactions and accretion. All three objects have enhanced star formation in their central regions (green contours indicate Hα emission) and the object to the left (Mrk 1477) clearly has a central star forming disk, a large polar ring and an extended HI tail or filament.

## 4. Future outlook

From the previous sections it is very clear that sensitive HI observations of galaxies provide a wealth of information about various processes that regulate the evolution of a galaxy. To date only the nearby universe has been probed at the required sensitivities. Surveys with existing instruments such as WHISP with the WSRT [27] and THINGS with the VLA [30] have provided a wealth of information assuring that probing the HI in galaxies provides essential and unique insights in the complex process of galaxy formation and evolution. New programmes are under way (e.g. the HALOGAS project with the WSRT [8] ) which push the sensitivity to their current limits in order to reveal the low surface brightness HI environment of some twenty galaxies and to detect the kind of faint HI signatures as described in section 2 and 3.

New instruments, in particular the SKA precursors and pathfinders such as ASKAP in Australia, MeerKAT in South Africa and APERTIF on the WSRT will provide information on HI in many more galaxies and many different environments. They will definitely contribute to a better under standing of galaxy evolution, in particular the role of the gas. The sensitivity will, however, not greatly increase over the current capabilities, so in order to detect the very low column density features as found in e.g. NGC 891 and NGC 6946, the collecting area of the full SKA will be required. Table 1 illustrates these capabilities for a range of distances (and corresponding resultions and HI mass sensitivities) in the nearby universe. It is clear that in order to discover the inportant HI signatures descrive here well beyond the Virgo cluster the impressive power of the SKA is requiered.

**Table 1.** HI detection limits and resolutions for the full SKA. For the evaluation of HI masses and column densities a velocity width of $\Delta v = 50$ km s$^{-1}$ was assumed. M(HI) and N(HI) have been determined per resolution element assuming a 12 hour integration.

| Longest Baseline (km) | Distance Mpc | Noise μJy | Resolution kpc | Resolution arcsec | M(HI) 5σ $M_\odot$ | N(HI) 1σ cm$^{-2}$ |
|---|---|---|---|---|---|---|
| 2 | 1 | 2.5 | 0.1 | 18.0 | $2.5 \times 10^2$ | $4.0 \times 10^{17}$ |
|  | 7 | 2.5 | 0.7 | 18.0 | $9.2 \times 10^3$ | $4.0 \times 10^{17}$ |
|  | 13 | 2.5 | 1.2 | 18.0 | $3.1 \times 10^4$ | $4.0 \times 10^{17}$ |
|  | 19 | 2.5 | 1.7 | 18.0 | $6.5 \times 10^4$ | $4.0 \times 10^{17}$ |
| 6 | 1 | 2.5 | 0.04 | 6.0 | $2.1 \times 10^2$ | $3.6 \times 10^{18}$ |
|  | 7 | 2.5 | 0.2 | 6.0 | $7.6 \times 10^3$ | $3.6 \times 10^{18}$ |
|  | 13 | 2.5 | 0.4 | 6.0 | $2.6 \times 10^4$ | $3.6 \times 10^{18}$ |
|  | 19 | 2.5 | 0.6 | 6.0 | $5.4 \times 10^4$ | $3.6 \times 10^{18}$ |
| 50 | 1 | 1.5 | 0.004 | 0.7 | $1.7 \times 10^2$ | $1.5 \times 10^{20}$ |
|  | 7 | 1.5 | 0.03 | 0.7 | $6.1 \times 10^3$ | $1.5 \times 10^{20}$ |
|  | 13 | 1.5 | 0.05 | 0.7 | $2.1 \times 10^4$ | $1.5 \times 10^{20}$ |
|  | 19 | 1.5 | 0.07 | 0.7 | $4.3 \times 10^4$ | $1.5 \times 10^{20}$ |